\documentclass[prl, pdflatex, twocolumn, superscriptaddress, notitlepage, floatfix]{revtex4-1}
\usepackage[toc,page,header]{appendix}
\usepackage{minitoc}

\usepackage{xcolor}
\usepackage{graphicx}
\usepackage{wrapfig}
\usepackage{amsmath}
\usepackage{amssymb}
\usepackage{multirow}
\usepackage{slashed}
\usepackage{physics}
\usepackage[colorlinks=true, linkcolor=blue, citecolor=blue, urlcolor=blue]{hyperref}
\usepackage{booktabs}
\usepackage{bbm}

\begin{document}
\doparttoc 
\faketableofcontents 

\part{} 

\title{Baryon electric charge correlation as a magnetometer of QCD}

\author{H.-T. Ding}
\address{Key Laboratory of Quark and Lepton Physics (MOE) and Institute of
Particle Physics, Central China Normal University, Wuhan 430079, China}
\author{J.-B. Gu}
\email{jinbiaogu@mails.ccnu.edu.cn}
\address{Key Laboratory of Quark and Lepton Physics (MOE) and Institute of
Particle Physics, Central China Normal University, Wuhan 430079, China}
\author{A. Kumar}
\address{Key Laboratory of Quark and Lepton Physics (MOE) and Institute of
Particle Physics, Central China Normal University, Wuhan 430079, China}
\author{S.-T. Li}
\affiliation{Key Laboratory of Quark and Lepton Physics (MOE) and Institute of
Particle Physics, Central China Normal University, Wuhan 430079, China}
\author{J.-H. Liu}
\affiliation{Key Laboratory of Quark and Lepton Physics (MOE) and Institute of
Particle Physics, Central China Normal University, Wuhan 430079, China}
\date{\today}% 

\begin{abstract}
The correlation between net baryon number and electric charge, $\chi_{11}^{\rm BQ}$, can serve as a magnetometer of QCD. This is demonstrated by lattice QCD computations using the highly improved staggered quarks with physical pion mass of $M_\pi=135~$MeV on $N_\tau=8$ and 12 lattices. We find that $\chi_{11}^{\rm BQ}$ along the transition line starts to increase rapidly with magnetic field strength $eB\gtrsim 2M_\pi^2$ and by a factor 2 at $eB\simeq 8M_\pi^2$. Furthermore, the ratio of electric charge chemical potential to baryon chemical potential, $\mu_{\rm Q}/\mu_{\rm B}$, shows significant dependence on the magnetic field strength and varies from the ratio of electric charge to baryon number in the colliding nuclei in heavy ion collisions. These results can provide baselines for effective theory and model studies, and both $\chi_{11}^{\rm BQ}$ and $\mu_{\rm Q}/\mu_{\rm B}$ could be useful probes for the detection of magnetic fields in relativistic heavy ion collision experiments as compared with corresponding results from the hadron resonance gas model.

\end{abstract}

\maketitle

\emph{Introduction.--}
Strong magnetic fields are expected to be created in various systems, including the early universe~\cite{Vachaspati:1991nm}, magnetars~\cite{enqvist1993primordial}, as well as in the laboratory of relativistic heavy ion collisions~\cite{Kharzeev:2007jp,Skokov:2009qp,Deng:2012pc}. In non-central relativistic heavy ion collisions, the strength of the produced magnetic field $eB$ can reach the order of $\Lambda_{\rm QCD}^2$, a typical scale of the strong interaction. Theoretical studies showed that the maximum magnetic field strength can reach $5M_\pi^2$ and $70M_\pi^2$ in Au+Au collisions at the top energy of Relativistic Heavy Ion Collisions (RHIC) experiments and in Pb+Pb collisions at Large Hadron Collider (LHC) energies~\cite{Skokov:2009qp,Deng:2012pc}, respectively, where $M_\pi$ is the mass of the lightest hadron, pion at vanishing magnetic fields. Thus, such a strong magnetic field can affect the phase structure of the strong interaction as described by Quantum Chromodynamics (QCD)~\cite{Kharzeev:2012ph}.

One of the most interesting effects induced by the strong magnetic field is the so-called chiral magnetic effect, which shows the macroscopic manifestation of the chiral anomaly of gauge fields. It was proposed in 2007 in the context of heavy ion collisions~\cite{Kharzeev:2007jp}, where strong magnetic field, at the order of $\Lambda_{\rm QCD}^2$, and axial U(1) anomaly are present. This has triggered intensive experimental as well as theoretical studies~\cite{Kharzeev:2020jxw}. However, results from searches for the chiral magnetic effect in heavy ion collision experiments turn out to be bewildering~\cite{STAR:2021mii,STAR:2022ahj,Kharzeev:2022hqz}, and it is only in the condensed matter experiments that evidence for the chiral magnetic effects has been found~\cite{Li:2014bha}. Among many different perspectives between these two kinds of experiments~\cite{Kharzeev:2020jxw,Li:2014bha}, one of the key differences is that the magnetic field is expected to decay fast in the former case while it is sustainable in the latter case.

Unfortunately, it is a challenging task to determine the lifetime of a magnetic field produced in the heavy ion collision experiments~\cite{Huang:2022qdn,Wang:2021oqq}. Theoretically, the lifetime depends on the electrical conductivity and types of magnetism of the medium. Recent first-principle lattice QCD studies have found that the electrical conductivity along the magnetic field increases as the magnetic field grows~\cite{Astrakhantsev:2019zkr}, and the quark-gluon plasma exhibits paramagnetic properties~\cite{Bali:2020bcn}. These two findings support the idea that the magnetic field could live longer in the evolution of heavy ion collisions than in the vacuum. Hints have been found for the manifestation of magnetic field in the deconfined quark-gluon plasma phase through recent observations of differences of direct flows between $D^0$ and $\bar{D}^0$~\cite{Adam:2019wnk,Acharya:2019ijj} and the broadening of transverse momentum distribution of dileptons produced through photon fusion processes~\cite{Adam:2018tdm,Aaboud:2018eph} in heavy ion collisions. On the other hand, thermal quantities such as chiral condensates, screening masses, and heavy quark potential have been found to be largely affected by the strong magnetic field via the first-principle lattice QCD studies~\cite{Bali:2012zg,Bonati:2016kxj,Bonati:2014ksa,DElia:2021tfb}. Unfortunately, these quantities are not directly measurable in the heavy ion collision experiments.

Among thermodynamic quantities accessible in both theoretical computations and experimental measurements, fluctuations of and correlations among net baryon number (B), electric charge (Q), and strangeness (S) are useful probes to study the changes in degrees of freedom and the QCD phase structure~\cite{Ding:2015ona,Fu:2022gou,Luo:2017faz,Pandav:2022xxx,Rustamov:2022hdi,Nonaka:2023xkg,Ko:2023eff}. However, they are much less explored at nonzero magnetic fields. Most of the studies have been carried out within the framework of hadron resonance gas 
(HRG) model~\cite{Fukushima:2016vix,Ferreira:2018pux,Bhattacharyya:2015pra,Kadam:2019rzo}, the Polyakov-Nambu-Jona-Lasinio model~\cite{Fu:2013ica}, and the Polyakov loop extended chiral
SU(3) quark mean field model~\cite{Chahal:2023khc}. The only existing lattice QCD study on the fluctuations of and correlations among conserved charges at nonzero magnetic fields was conducted using the larger than physical pion mass at one single lattice cutoff~\cite{Ding:2021cwv}.

In this letter, we present the first lattice QCD computation with physical pion mass on the quadratic fluctuations and correlations of net baryon number, electric charge, and strangeness in the presence of constant external magnetic fields. Both the correlation among baryon number and electric charge, $\chi_{11}^{\rm BQ}$, and the ratio of electric charge chemical potential over baryon chemical potential, $\mu_{\rm Q}/\mu_{\rm B}$, are found to be significantly enhanced in the magnetic field and could be useful to detect the existence of magnetic field in heavy ion collision experiments. Some of the preliminary results are presented in~\cite{Ding:2022uwj}.

\emph{Quadratic fluctuations of conserved charges and the HRG model in strong magnetic fields--}
The quadratic fluctuations of and correlations among B, Q, and S can be obtained by taking the derivatives of pressure with respect to the chemical potentials $\hat{\mu}_X\equiv \mu_X/T$ with $X=$ B, Q, and S from lattice calculation evaluated at zero chemical potentials
\begin{equation}
\begin{aligned}
{\chi}_{i j k}^{\mathrm{BQS}} &=\left.\frac{\partial^{i+j+k} P / T^{4}}{\partial\hat{\mu}_{\mathrm{B}}^{i} \partial\hat{\mu}_{\mathrm{Q}}^{j} \partial\hat{\mu}_{\mathrm{S}}^{k}}\right|_{\hat{\mu}_{\mathrm{B}, \mathrm{Q}, \mathrm{S}}=0},
\end{aligned}
\label{eq:sus}
\end{equation}
where $P =\frac{T}{V} \ln Z(eB,\mu_{\rm B},\mu_{\rm Q},\mu_{\rm S})$ denotes the total pressure of the hot magnetized medium, and $i+j+k=2$. For brevity, we drop the superscript when the corresponding subscript is zero.

In the context of the HRG model, the thermal pressure in strong magnetic fields arising from charged hadrons can be expressed as follows~\cite{Ding:2021cwv,Fukushima:2016vix,Endrodi:2013cs}

\begin{align}
    \frac{P_c}{T^4} =& \frac{|q_i| B}{2\pi^2T^3} \sum_{s_z=-s_i}^{s_i} \sum_{l=0}^\infty \varepsilon_0 \sum_{n=1}^\infty (\pm 1)^{n+1} \frac{e^{n\mu_i/T}}{n} {\rm K}_1 \left(\frac{n\varepsilon_0}{T}\right) ,
    \label{eq:HRGp}
\end{align}
where $\varepsilon_0$ is the energy level of charged hadrons and has a form of $\varepsilon_{0}=\sqrt{m_{i}^{2}+2\left|q_{i}\right| B\left(l+1 / 2-s_{z}\right)}$.
Here $q_{i}$ and $m_{i}$ are the electric charge and mass of the hadron $i$, respectively, while $s_z$ is the spin factor which is summed over $-s_i$ to $s_i$ for each hadron $i$. $B$ is the magnetic field pointing along the $z$ direction, and $l$ denotes the Landau levels. $n$ is the sum index in the Taylor expansion series and $K_1$ is the first-order modified Bessel function of the second kind. The “$+$” in “$\pm$” corresponds to the case for mesons ($s_{i}$ is integer) while the “$-$” for baryons ($s_{i}$ is half-integer). Note that the HRG description of spin-3/2 baryons as well as spin-1 mesons breaks down at some critical magnetic field, above which the lowest energy of the particle would turn negative. In our case, the largest $eB$ applied is $\sim8M_\pi^2$, which keeps $\varepsilon_0$ always positive.

The quadratic fluctuations of and correlations among B, Q, and S arising from charged hadrons, are thus given by~\cite{Ding:2021cwv}~\footnote{Note that~\autoref{eq:HRGsus} is obtained from thermal pressure ~\autoref{eq:HRGp} and its derivatives with respect to chemical potentials (cf.~\autoref{eq:sus}), exploiting the fact that the vacuum pressure, although dependent on $eB$, is independent of chemical potentials~\cite{Fukushima:2016vix,Ding:2021cwv,Endrodi:2013cs}.}
\begin{equation}
    \begin{aligned}
    &{\chi}_{2}^{X}=\frac{B}{2 \pi^{2} T^3} \sum_{i}\left|q_{i}\right| X_{i}^{2} \sum_{s_{z}=-s_{i}}^{s_{i}} \sum_{l=0}^{\infty} f\left(\varepsilon_{0}\right) \,,\\
    &{\chi}_{11}^{X Y}=\frac{B}{2 \pi^{2} T^3} \sum_{i}\left|q_{i}\right| X_{i} Y_{i} \sum_{s_{z}=-s_{i}}^{s_{i}} \sum_{l=0}^{\infty} f\left(\varepsilon_{0}\right)\,,
    \end{aligned}
    \label{eq:HRGsus}
\end{equation}
where $f\left(\varepsilon_{0}\right)=\varepsilon_{0} \sum_{n=1}^{\infty}(\pm 1)^{n+1} n \mathrm{~K}_{1}\left(\frac{n \varepsilon_{0}}{T}\right)$ and $X_i,Y_i=$ B, Q, S carried by hadron $i$. The fluctuations and correlations arising from neutral hadrons are obtained using the standard HRG model~\cite{HotQCD:2012fhj,Bollweg:2021vqf} as the masses of neutral hadrons are assumed to be independent on $eB$ in the current $eB$ window. In the current HRG model computations, we adopt the list of resonances from QMHRG2020~\cite{Bollweg:2021vqf}.

\begin{figure}[!htp]
\begin{center}
    \includegraphics[width=0.38\textwidth]{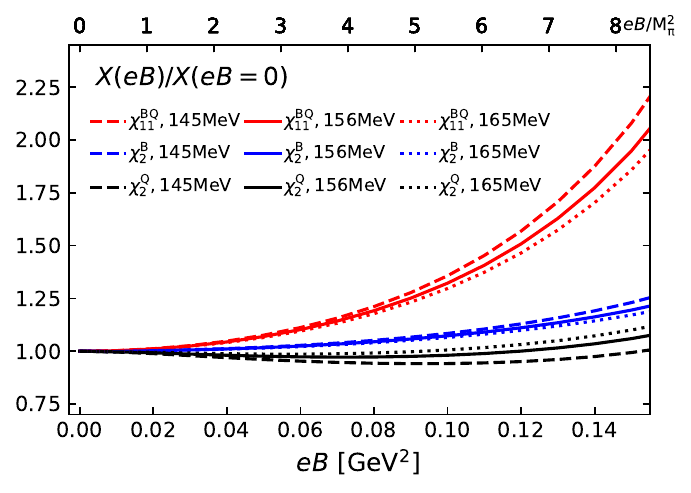}
    \caption{The ratio of $X=\chi_{11}^{\rm BQ},\chi_{2}^{\rm B}$, and $\chi_{2}^{\rm Q}$ to its corresponding value at vanishing magnetic fields as a function of $eB$ at three temperatures obtained from the HRG model.} 
    \label{fig:HRG}		
\end{center}
\end{figure}

\autoref{fig:HRG} shows the $eB$ dependence of normalized $\chi_{11}^{\rm BQ}$, $\chi_{2}^{\rm B}$, and $\chi_{2}^{\rm Q}$ obtained from the HRG model. Note that both $\chi_{11}^{\rm BQ}$ and $\chi_2^{\rm Q}$ receive contributions only from charged hadrons, while $\chi_2^{\rm B}$ receives contributions from both charged and neutral baryons. It can be seen that $\chi_{11}^{\rm BQ}$ increases rapidly as $eB$ grows and reaches a factor of $\sim 1.9$ at $eB\simeq 8M_\pi^2$. On the other hand, $\chi_{2}^{\rm B}$ has much weaker dependence on $eB$ and increases about 20\% while $\chi_{2}^{\rm Q}$ remains almost intact as $eB$ grows.

\begin{figure*}[!htp]	
\begin{center}
    \includegraphics[width=0.32\textwidth]{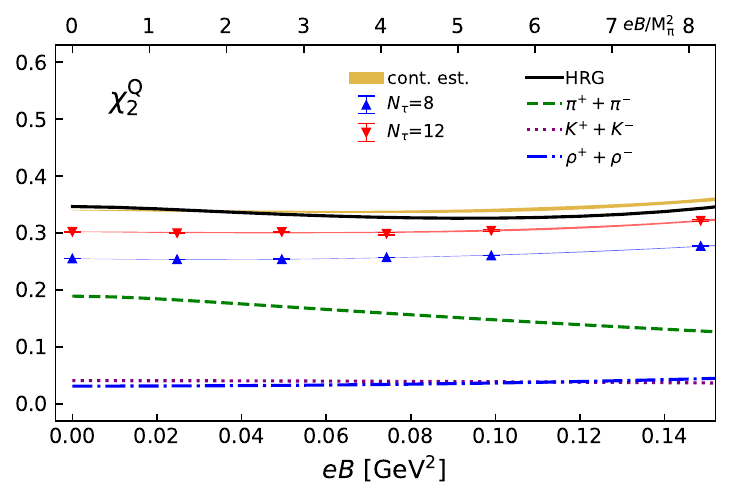}
    \includegraphics[width=0.32\textwidth]
    {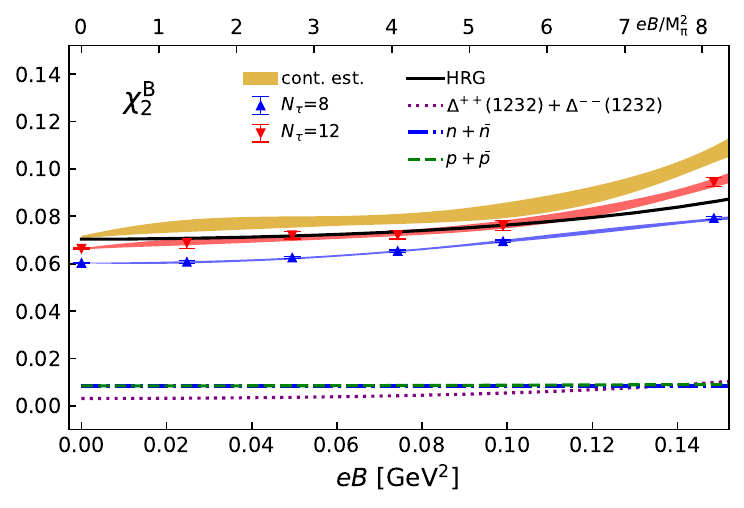} 
    \includegraphics[width=0.32\textwidth]{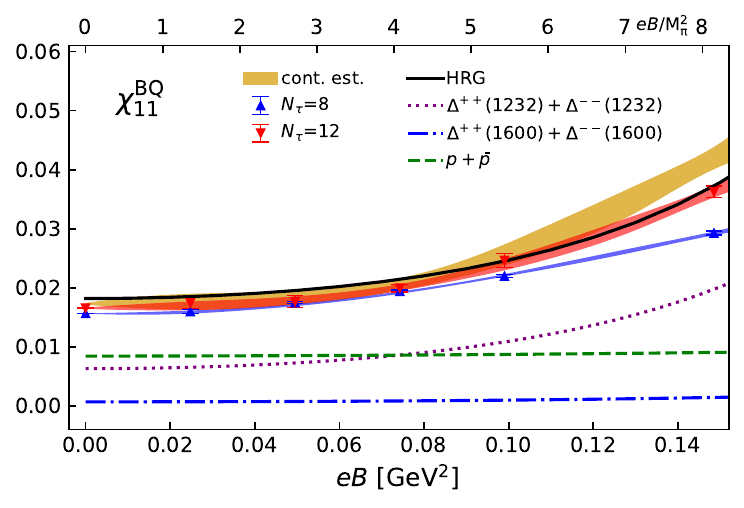}
    \caption{Continuum estimates (yellow bands) of  $\chi_2^{\rm Q}$ (left), $\chi_{2}^{\rm B}$ (middle), and $\chi_{11}^{\rm BQ}$ (right) at $T=145$ MeV based on lattice QCD results with $N_\tau=8$ and $12$. The blue and red bands represent the interpolated results for $N_\tau=8$ and $12$ lattice data, respectively. The total contribution (black solid lines) as well as contributions from certain hadrons (broken lines) to $\chi_2^{\rm Q}$, $\chi_{2}^{\rm B}$, and $\chi_{11}^{\rm BQ}$ obtained from the HRG model are also shown.}
    \label{fig:QsusbelowTc}		
\end{center}
\end{figure*}

\emph{Lattice QCD simulations--}The partition function $Z$ of QCD with three flavors ($f=u, d, s$) is given by the functional integral,
\begin{align}
Z=\int \mathcal{D}U e^{-S_g}\prod_{f=u,d,s}[\det M(U,q_fB,\mu_f)]^{\frac{1}{4}}\,.
\end{align} 
The highly improved staggered quarks (HISQ)~\cite{Follana:2006rc} and a tree-level improved Symanzik gauge action, which have been extensively used by the HotQCD collaboration~\cite{Bazavov:2011nk,HotQCD:2014kol,HotQCD:2012fhj,Bazavov:2017dus, Bazavov:2018mes,Bazavov:2019www,HotQCD:2019xnw,Bollweg:2021vqf}, were adopted in our current lattice simulations of $N_f=2+1$ QCD in nonzero magnetic fields on $32^3\times8$ and $48^3\times12$ lattices. The magnetic field is introduced along the $z$ direction and described by a fixed factor $u_{\mu}(n)$ of the U(1) field.
We set the quark masses to their physical values, with mass degenerate light quarks $m_u=m_d$ corresponding to $M_\pi=135$ MeV. The electric
charges of the quarks are $q_d=q_s=-q_u/2=-e/3$, with $e$ denoting the elementary electric charge.
To satisfy the quantization for all the quarks in the system, the greatest common divisor of their electric charges, i.e., $\left|q_{d}\right|=\left|q_{s}\right|=e /3$, is adopted, and the strength of the magnetic field $e B$ thus equals to $\frac{6 \pi N_{b}}{N_{x} N_{y}} a^{-2}$~\cite{DElia:2010abb,Bali:2011qj}. Here $N_{b}$ is the number of magnetic fluxes through a unit area in the x-y plane, $a$ is the lattice spacing, and $N_\sigma\equiv N_x=N_y$ are the spatial lattice points. Further details on the implementation of magnetic fields in the lattice QCD simulations using the HISQ action can be found in~\cite{Ding:2020hxw}. The simulated $eB$ ranges from $\simeq M_\pi^2$ up to $\simeq 8M_\pi^2$, with $N_b$ varying from 1 to 6. The discretization error in $eB$ should be mild as $N_b/N_\sigma^2\ll 1$~\cite{Endrodi:2019zrl}.

All gauge configurations have been generated using a modified version of the software suite \texttt{SIMULATeQCD}~\cite{HotQCD:2023ghu} and saved every tenth time units. For each value of $N_b$ at temperatures below $160$ MeV, about 40,000 configurations were saved on $N_\tau=8$ lattices and 5,000 on $N_\tau=12$ lattices. Approximately 3000 configurations are additionally generated on $N_\tau=16$ lattices at a single temperature with $N_b=3$ to scrutinize the uncertainties originating from continuum estimates. Fluctuations and correlations of conserved charges up to the 4th order in nonzero magnetic fields have been computed using the random noise vector method. Details of computations can be found in ~\autoref{tab:stat} in the Supplemental Materials. For the case of $N_b=0$, we adopted lattice QCD results obtained in~\cite{Bollweg:2021vqf}.

\emph{Results--} We first present in~\autoref{fig:QsusbelowTc} the results for $\chi_2^{\rm Q}$, $\chi_{2}^{\rm B}$, and $\chi_{11}^{\rm BQ}$, as obtained from lattice QCD computations and the HRG model at ${\rm T}=145$ MeV, which is below the transition temperature $\sim 156$ MeV~\footnote{The transition temperature is roughly independent of $eB$ within the current $eB$ window. This is determined by the chiral susceptibility, as detailed in the Supplemental Materials}. The lattice QCD results are continuum estimated based on $N_\tau=8$ and 12 lattices, and the details are presented in the Supplemental Materials, where~\autoref{fig:cont_BQ_q1_eB} confirms the consistency between continuum estimated and extrapolated results, implying minor uncertainties. It can be seen that the continuum estimated lattice QCD result of $\chi_2^{\rm Q}$ remains almost intact with $eB$, whereas $\chi_2^{\rm B}$ increases by about 45\% at $eB\simeq8M_\pi^2$. Most strikingly, $\chi_{11}^{\rm BQ}$ is significantly affected by the magnetic field, increasing by a factor of $\sim 2.4$ at $eB\simeq8M_\pi^2$. On the other hand, while the HRG model provides a reasonable description of $\chi_2^{\rm Q}$ and $\chi_{11}^{\rm BQ}$ for field strengths up to $eB\lesssim 4 M_\pi^2$, it begins to undershoot continuum estimated lattice QCD results at higher $eB$ values. For $\chi_2^{\rm B}$, the HRG model undershoots the QCD results across the whole $eB$ window.

We break down the contributions from individual hadrons in the HRG model. In the case of $\chi_{2}^{\rm Q}$, the dominant contribution in the current window of magnetic field strength always comes from charged pions, although it decreases by about 30\% at $eB\simeq 8M_\pi^2$ due to their enhanced energy in the magnetic field. In the case of $\chi_{2}^{\rm B}$, no single hadron overwhelmingly dominates; the largest contribution, which is less than 12\%, comes from protons.
For $\chi_{11}^{\rm BQ}$, it can be seen that protons dominate the contribution at $eB\lesssim 4M_\pi^2$, while doubly charged $\Delta(1232)$ baryons start to surpass protons at $eB\gtrsim 4M_\pi^2$. Note that the contribution from protons almost remains constant with $eB$. Thus, most of the $eB$ dependence of $\chi_{11}^{\rm BQ}$ comes from doubly charged $\Delta(1232)$ baryons. This follows from the fact that the energy $\epsilon_0$ of $\Delta(1232)$ baryons can become smaller, as they are doubly charged and have a spin of $3/2$. For other doubly charged baryons, e.g., $\Delta(1600)$, their contributions are largely suppressed due to their larger masses.

\begin{figure}[!htp]			
    \includegraphics[width=0.35\textwidth]{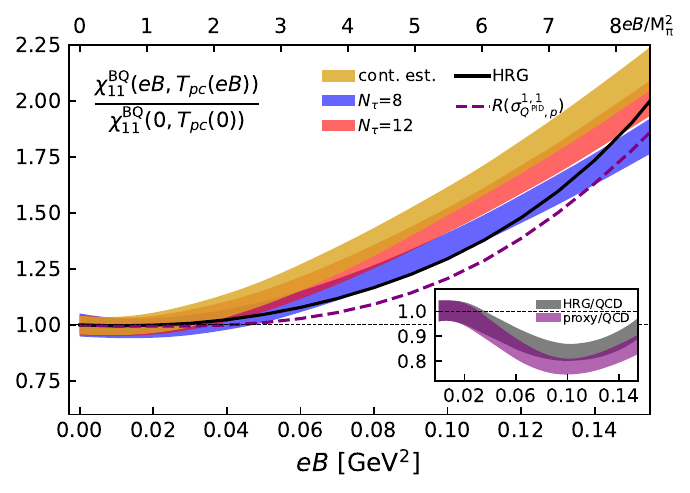}
    \includegraphics[width=0.35\textwidth]{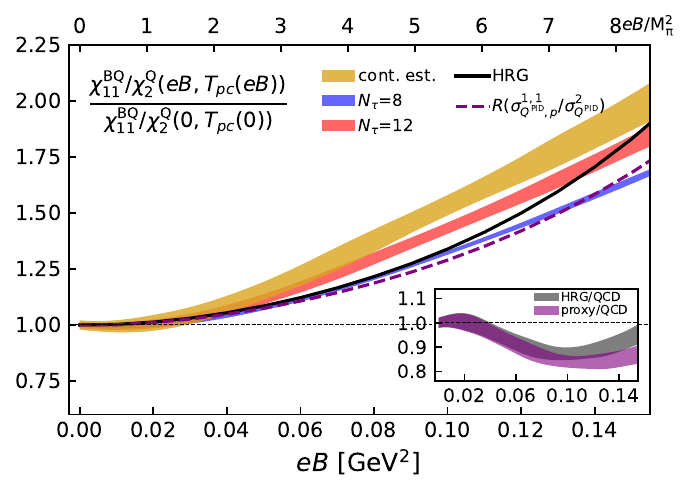}	
    \caption{Continuum estimates of ratios of $\chi_{11}^{\rm BQ}$ (top) and $\chi_{11}^{\rm BQ}/\chi_{2}^{\rm Q}$ (bottom) to their corresponding values at vanishing magnetic fields along the transition line. Interpolated bands for $N_\tau=8$ and 12 lattice data are also shown. The insets show ratios of results obtained from the HRG model and proxy to continuum estimated lattice QCD results. }
    \label{fig:ratioTpc}
\end{figure}

In heavy ion collisions, correlations among conserved charges are measured through final stable particles. For instance, the proton ($p$) serves as a proxy for baryon number, and net electric charge ($Q^{\rm PID}$) is measured through proton, pion, and kaon~\cite{STAR:2019ans}. However, the doubly charged $\Delta(1232)$ baryons, which significantly contribute to the $eB$ dependence of $\chi_{11}^{\rm BQ}$, are not directly measurable in heavy ion collision experiments. This is because they are short-lived resonances, undergoing a strong decay into proton and pion with a branching ratio close to 100\%. To determine whether the decays of $\Delta(1232)$ baryons, i.e. protons and pions, retain the memory of the $eB$ dependence of $\Delta(1232)$'s contribution to $\chi_{11}^{\rm BQ}$, we construct a proxy $\sigma_{Q^{\rm PID},p}^{1,1}$ for $\chi_{11}^{\rm BQ}$ that includes contributions from all the decays to proton and pion following the standard approach in the framework of the HRG model~\footnote{In our computation of $\sigma_{Q^{\rm PID},p}^{1,1}$ we assume that decay channels and relevant branching ratios in the nonzero magnetic fields remain the same as those in the vacuum, i.e. following the procedure outlined in~\cite{Bellwied:2019pxh} for zero magnetic fields. More information is provided in Section IV of Supplemental Materials.}.

It is common to investigate the ratios of fluctuations in both theory and experiments to suppress the dependence on the system volume~\cite{Ding:2015ona,Fu:2022gou,Luo:2017faz,Pandav:2022xxx,Rustamov:2022hdi}. We then focus on ratios $R(\mathcal{O})\equiv\mathcal{O}(eB,T_{pc}(eB))/\mathcal{O}(0,T_{pc}(0))$ along the transition line. In \autoref{fig:ratioTpc} (top), the continuum estimate of $R(\chi_{11}^{\rm BQ})$ at $T_{pc}(eB)$ exhibits a significant $eB$ dependence, similar to observations in \autoref{fig:QsusbelowTc} at ${\rm T}=145$ MeV. At the highest $eB$ value, $\sim8 M_\pi^2$, $R(\chi_{11}^{\rm BQ})$ reaches about 2.1. For $eB\lesssim M_\pi^2$, both the proxy $R(\sigma_{Q^{\rm PID},p}^{1,1})$ (dashed line) and the HRG result (solid line) are consistent with the continuum estimated lattice QCD result. However, for $eB\gtrsim M_\pi^2$, both $R(\sigma_{Q^{\rm PID},p}^{1,1})$ and HRG results begin to undershoot the continuum estimated lattice QCD result. As depicted in the inset, the proxy underestimates the continuum estimated lattice QCD result by $\sim$22\% at most at $eB\simeq 5.5 M_\pi^2$ and by $\sim$16\% at $eB\simeq 8 M_\pi^2$. In the case of HRG, this underestimation is by $\sim$15\% at most at $eB\simeq 5.5 M_\pi^2$ and by $\sim$9\% at $eB\simeq 8 M_\pi^2$.
In \autoref{fig:ratioTpc} (bottom), a similar trend is observed in the double ratio, $R(\chi_{11}^{\rm BQ}/\chi_{2}^{\rm Q})$, where the proxy $R(\sigma_{Q^{\rm PID},p}^{1,1}/\sigma_{Q^{\rm PID}}^2)$ provides a slightly better description of the continuum estimated lattice QCD result compared to the case of $R(\chi_{11}^{\rm BQ})$.

Furthermore, the electric charge chemical potential can be expanded as ${\hat{\mu}_{\rm Q}}=q_1\hat{\mu}_{\rm B} + q_3 \hat{\mu}_{\rm B}^3+\mathcal{O}(\hat{\mu}_{\rm B}^5)$. Since the initial nuclei in heavy ion collisions are net strangeness neutral, the leading order coefficient $q_1$ can be expressed as follows~\cite{Bazavov:2012vg,Bazavov:2014xya}
\begin{equation} 
    q_1 =
    \frac{
    r\left( \chi_2^{\rm B}\chi_2^{\rm S} - \chi_{11}^{\rm BS}\chi_{11}^{\rm BS} \right)
    -\left( \chi_{11}^{\rm BQ}\chi_2^{\rm S} -\chi_{11}^{\rm BS} \chi_{11}^{\rm QS} \right)
    }{
    \left( \chi_2^{\rm Q}\chi_2^{\rm S}  - \chi_{11}^{\rm QS} \chi_{11}^{\rm QS} \right)
    - r \left(\chi_{11}^{\rm BQ}\chi_2^{\rm S} - \chi_{11}^{\rm BS}\chi_{11}^{\rm QS} \right)
    }.
    \label{eq:muQbymuB}
\end{equation}
Here $r \equiv {n_{\rm Q}}/{n_{\rm B}}$ stands for the ratio of net electric charge to net baryon number density in the colliding nuclei. For $\rm{Au}$+$\rm {Au}$ and Pb+Pb collisions, $r=0.4$ is a suitable approximation. In the case of isobar collisions, for $_{40}^{96}\rm{Zr}$+$_{40}^{96}\rm{Zr}$, $r$ is marginally higher at $r=0.417$, while for $_{44}^{96}\rm{Ru}$+$_{44}^{96}\rm{Ru}$, $r$ is 10\% larger, specifically $r=0.458$.

\begin{figure}[h!]			
    \includegraphics[width=0.35\textwidth]{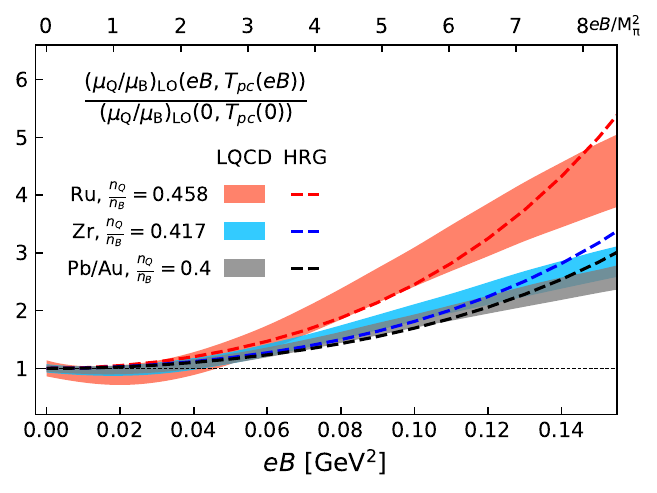} 
    \caption{The continuum estimated $(\mu_{\rm Q}/\mu_{\rm B})_{\rm LO}$ normalized to its  value at $eB=0$ as a function of $eB$ along the transition line. Bands correspond to collision systems with various values of $n_{\rm Q}/n_{\rm B}$ and lines are corresponding results obtained from the HRG model.}
    \label{fig:muQbymuB}
\end{figure}

In~\autoref{fig:muQbymuB}, we show the leading order contribution to $\mu_{\rm Q}/\mu_{\rm B}$, denoted as $(\mu_{\rm Q}/\mu_{\rm B})_{\rm LO}$, normalized to its value at zero magnetic fields as a function of $eB$ along the transition line for various values of $r$ corresponding to different collision systems. It can be seen that the double ratio $R((\mu_{\rm Q}/\mu_{\rm B})_{\rm LO})$ increases with $eB$ across all collision systems. In Au+Au and Pb+Pb collisions, $R((\mu_{\rm Q}/\mu_{\rm B})_{\rm LO})$ reaches approximately 2.4 at $eB\simeq 8 M_\pi^2$. For the isobar collision systems, designed to study the chiral magnetic effect, $R((\mu_{\rm Q}/\mu_{\rm B})_{\rm LO})$ in  Zr+Zr collisions is comparable to that of Au+Au and Pb+Pb collisions. However, in Ru+Ru collisions, $R((\mu_{\rm Q}/\mu_{\rm B})_{\rm LO})$ increases more rapidly, reaching about 4 at $eB\simeq 8 M_\pi^2$, which is about 1.5 times greater compared to the other three cases. Additionally, we find that the contribution from the next-to-leading order term $q_3$, obtained on $N_\tau=8$ lattices, is about 2\% of that from the leading order, as detailed in the Supplemental Materials. The results obtained from the HRG model (denoted by the broken lines) exhibit reasonably good agreement with the lattice QCD data. This suggests that the observation of $eB$ dependence of $\mu_{\rm Q}/\mu_{\rm B}$ through fits to particle yields using the HRG model with magnetized hadron spectrum is feasible.

\emph{Conclusions.---} 
We have performed the first lattice QCD computations of quadratic fluctuations and correlations of conserved charges in nonzero magnetic fields with physical pions.
Based on these computations, we propose two probes to detect the imprints of magnetic fields in the final stages of heavy ion collisions: the second-order correlation of baryon number and electric charge ($\chi_{11}^{\rm BQ}$), and the ratio of electric charge chemical potential to baryon number chemical potential ($\mu_{\rm Q}/\mu_{\rm B}$).

Possible experimental analyses could be carried out across various centrality classes or in different collision systems exhibiting distinct $eB$ values, as the strength of magnetic fields is expected to increase from central to peripheral collisions and in collisions of isobars with larger number of protons~\cite{Skokov:2009qp,Deng:2012pc}.  The $\chi_{11}^{\rm BQ}$ could be investigated using its proxy~\cite{STAR:2019ans}, i.e. the proxy of $\chi_{11}^{\rm BQ}$ or $\chi_{\rm  11}^{\rm  BQ}/\chi_2^{\rm Q}$ can be obtained in the experiments by measuring the net proton number as the net baryon number ${\rm B}$, and the net proton, pion and kaon number as the net electric charge ${\rm Q}$, cf. $R(\sigma_{Q^{^{\rm PID}},p}^{1,1})$ and $R(\sigma_{Q^{^{\rm PID}},p}^{1,1}/\sigma_{Q^{\rm PID}}^2)$ as shown in~\autoref{fig:ratioTpc}. On the other hand, the ratio $\mu_{\rm Q}/\mu_{\rm B}$ can be obtained from thermal fits to particle yields~\cite{Braun-Munzinger:2003pwq,STAR:2017sal,Wheaton:2004qb}, employing the HRG model with magnetized hadron spectrum. Here, in addition to the normal free parameter $\mu_{\rm Q}$, $\mu_{\rm B}$ and temperature, an additional parameter $eB$ is needed in the thermal fits to accommodate the change in the hadron mass~\footnote{The HRG already gives a reasonably good description of the lattice data, hence, the inclusion of feeddown effects in the thermal fits may not be necessary.}. Moreover, it is worth noting that the strict normalization of these quantities to the case with $eB=0$, corresponding to the most central collision, may not be essential. Instead, one can directly investigate the dependence of $\chi_{11}^{\rm BQ}/\chi_2^{\rm Q}$ and $\mu_{\rm Q}/\mu_{\rm B}$ on centrality class and collision systems.
These analyses can utilize already existing data from facilities at LHC and RHIC~\cite{Pandav:2022xxx,Rustamov:2022hdi,Nonaka:2023xkg,Ko:2023eff}.

Finally, our results also establish QCD baselines in external magnetic fields for effective theories and model studies~\cite{Deng:2012pc,Huang:2022qdn,
Fu:2013ica,Fukushima:2016vix,Ferreira:2018pux,Bhattacharyya:2015pra,Kadam:2019rzo,Chahal:2023khc}, providing valuable insights into the dynamic evolution within heavy ion collisions.

We thank Xiaofeng Luo, Swagato Mukherjee, and Nu Xu for useful discussions.
This material is based upon work supported partly by the National Natural Science Foundation of China under Grants No. 12293064, No. 12293060, and No. 12325508, as well as the National Key Research and Development Program of China under Contract No. 2022YFA1604900. The numerical simulations have been performed on the GPU cluster in the Nuclear Science Computing Center at Central China Normal University ($\mathrm{NSC}^{3}$) and Wuhan Supercomputing Center. 

\bibliographystyle{apsrev4-1.bst}
\bibliography{refs.bib}

\begin{widetext}
\newcounter{totalequations}
\section{Supplemental Materials}
We provide supplemental materials in the sequence according to the contents of the main material.

\section{I. Parameters used in lattice QCD simulations}
\label{sup:statistics}

The parameters for lattice QCD simulations on $N_{\tau}$ = 8, 12 and 16 lattices are summarized in~\autoref{tab:stat}. In this table $N_{\rm v1}$ and $N_{\rm v2}$ represent the number of random noise vectors used to calculate $D_1=\partial \mathrm{ln~det} M_f/\partial \mu_f|_{\mu_f=0}$ and $D_2=\partial^2 \mathrm{ln~det} M_f/\partial \mu_f^2|_{\mu_f=0}$, respectively, where $M_f$ is the staggered fermion matrix for quark flavor $f$ and $\mu_f$ is the corresponding flavor chemical potential. $N_{\rm v3,4}$ is the number of random noise vectors used in calculating third-order and fourth-order operators, $D_3=\partial^3 \mathrm{ln~det} M_f/\partial \mu_f^3|_{\mu_f=0}$ and $D_4=\partial^4 \mathrm{ln~det} M_f/\partial \mu_f^4|_{\mu_f=0}$, respectively.

\begin{table}[!htp]
    \centering
        \begin{tabular}{*{13}{c}}
        \toprule \hline \hline
        \toprule  
        \multirow{2}*{$N_\sigma^3\times N_\tau$}  
        &	\multirow{2}*{$T$ [MeV]}  
        &	\multirow{2}*{$\beta$}	 	
        &   \multirow{2}*{$am_l$} 
        &   \multirow{2}*{$am_s$} 
        &   \multicolumn{5}{c}{\# conf.}
        &   \multirow{2}*{$N_{\rm v1}$} 
        &   \multirow{2}*{$N_{\rm v2}$} 
        &   \multirow{2}*{$N_{\rm v3,4}$} 
        \\
        \cline{6-10} 
        &         &         &        &        & $N_b=1$ & $N_b=2$ & $N_b=3$ & $N_b=4$ & $N_b=6$ &       &      &      \\
        \midrule \hline 
                        &144.95  &  6.315  & 0.00281 & 0.0759&  42792  &  46280  &  38755  &  43228  &  46137  &  603  & 603  & 102  \\
                        &151.00  &  6.354  & 0.00270 & 0.0728&  39124  &  44223  &  48032  &  44180  &  43991  &  603  & 603  & 102  \\
        $32^3 \times 8$ &156.78  &  6.390  & 0.00257 & 0.0694&  51145  &  46919  &  42544  &  45424  &  44712  &  603  & 603  & 102  \\
                        &162.25  &  6.423  & 0.00248 & 0.0670&  26259  &  24041  &  26619  &  26214  &  32426  &  603  & 603  & 102  \\
                        &165.98  &  6.445  & 0.00241 & 0.0652&  25080  &  20450  &  23104  &  22048  &  24600  &  603  & 603  & 102  \\
                        \hline
                        &144.97  &  6.712  & 0.00181 & 0.0490&  5583   &  5335   &  5326   &  5370   &  5170   &  705  & 102  & -\\
                        &151.09  &  6.754  & 0.00173 & 0.0468&  5299   &  5098   &  5130   &  5033   &  5160   &  606  & 102  & -\\
        $48^3 \times 12$&157.13  &  6.794  & 0.00167 & 0.0450&  4315   &  4464   &  4161   &  4241   &  4108   &  705  & 102  & -\\
                        &161.94  &  6.825  & 0.00161 & 0.0436&  5871   &  2456   &  2820   &  7318   &  4577   &  405  & 102  & -\\
                        &165.91  &  6.850  & 0.00157 & 0.0424&  3000   &  3000   &  2560   &  3000   &  2271   &  102  & 102  & -\\
                        \hline
        $64^3 \times 16$&156.92  &  7.095  & 0.00124 & 0.0334&     -   &     -   &  3052   &     -   &     -   &  603  & 102  & -\\
        \midrule
        \bottomrule \hline \hline	
        \end{tabular}
    \caption{Simulation parameters and statistics on $32^3 \times 8$, $48^3 \times 12$ and $64^3 \times 16$ lattices with light to strange quark mass ratio $m_l/m_s =1/27$. }
    \label{tab:stat}
\end{table}

\section{II. Two-dimensional spline fits in the $T-eB$ plane and continuum estimates}
\begin{figure}[!h]    
    \includegraphics[width=0.39\textwidth]{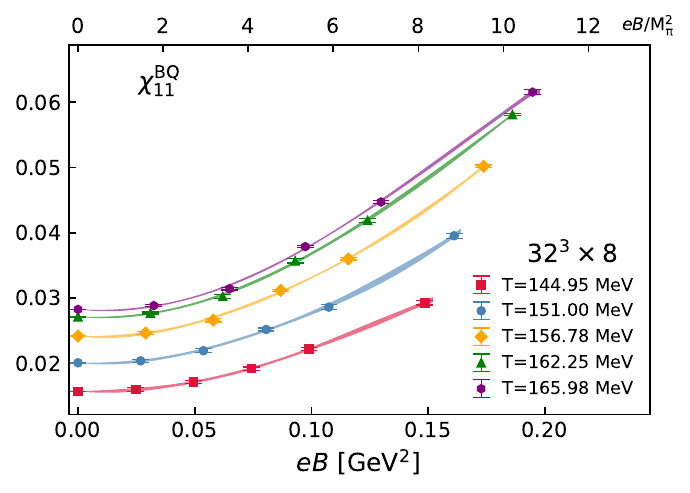}
    \includegraphics[width=0.39\textwidth]{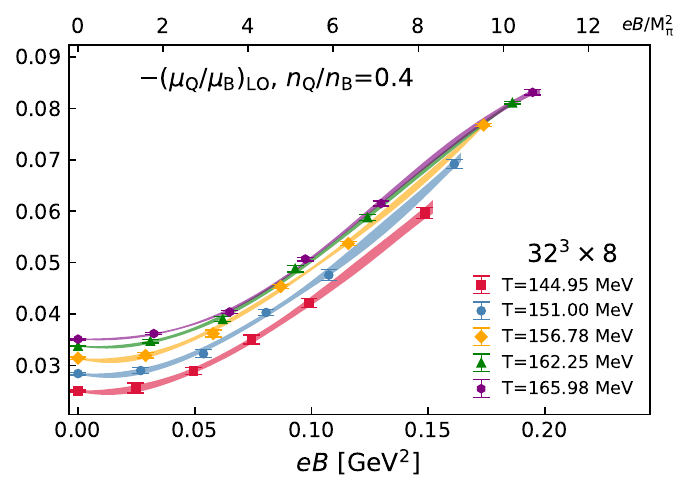}
    \includegraphics[width=0.39\textwidth]{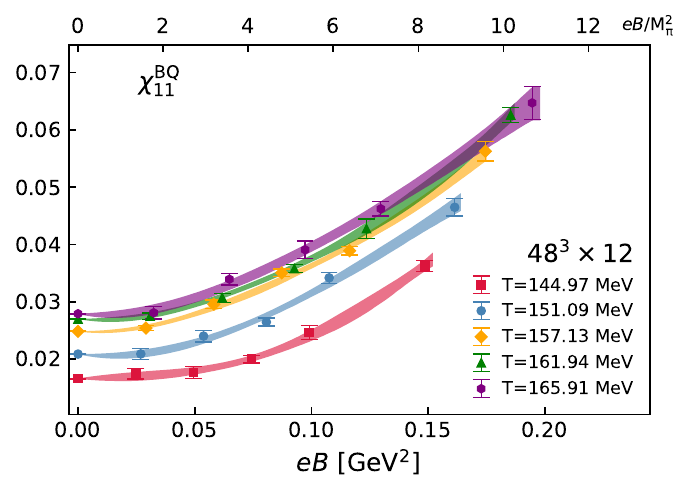}
    \includegraphics[width=0.39\textwidth]{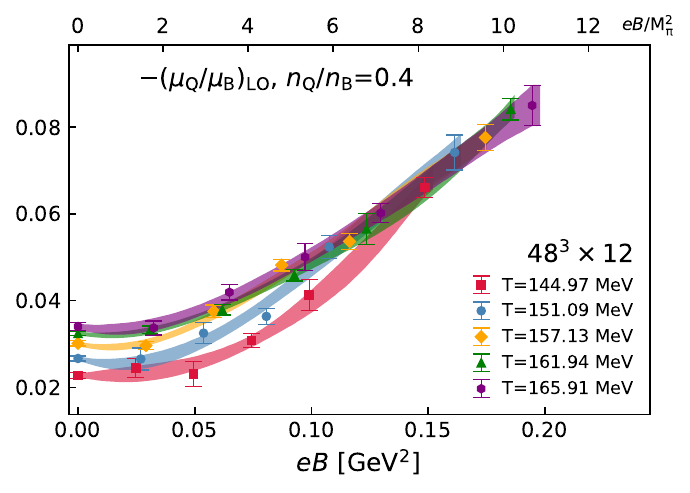}
    \caption{Spline fits to lattice data of $\chi_{11}^{\rm BQ}$ (left) and $-(\mu_{\rm Q}/\mu_{\rm B})_{\rm LO}$ with $n_{\rm Q}/n_{\rm B}=0.4$ (right). The top and bottom panels are for $N_\tau=8$ and $N_\tau=12$ lattices, respectively. The data points are obtained from lattice QCD computations, and the bands denote the two-dimensional B-spline fit results. The data points at $eB=0$ are taken from Ref.~\cite{Bollweg:2021vqf}.} 
    \label{fig:ob_nt8_12}
\end{figure}
Since external magnetic fields are quantized, meaning the number of magnetic fluxes denoted by $N_b$ does not vary continuously, interpolation in the $T-eB$ plane becomes essential. We adopt an approach similar to that in~\cite{Bali:2011qj} and in practice choose to use a two-dimensional B-spline function to fit our data. This allows us to deduce the $eB$ and $T$ dependence of targeted observables. In our analyses, error bands are obtained using the Gaussian bootstrapping and by performing spline fits on each sample. The final values and errors are taken as the median and 68\% percentiles of the bootstrap distribution.~\autoref{fig:ob_nt8_12} illustrates the spline fits to lattice data of $\chi_{11}^{\rm BQ}$ and $-(\mu_{\rm Q}/\mu_{\rm B})_{\rm LO}$ obtained on $N_\tau$ = 8 and $N_\tau$ = 12 lattices.

\begin{figure}[!h]    
    \centering           
    \includegraphics[width=0.39\textwidth]{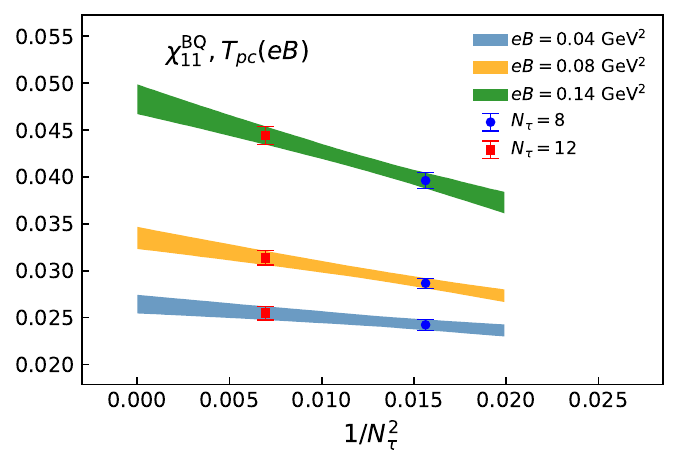}
    \includegraphics[width=0.39\textwidth]{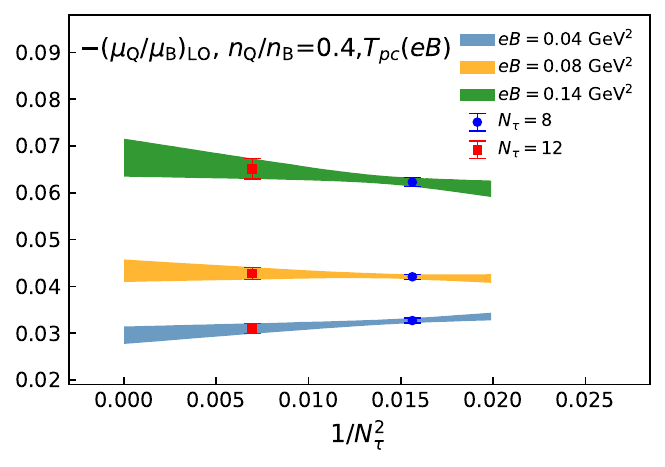}
    \caption{Continuum estimates of $\chi_{11}^{\rm BQ}$ (left) and $-(\mu_{\rm Q}/\mu_{\rm B})_{\rm LO}$ with $n_{\rm Q}/n_{\rm B}=0.4$ (right) along the transition line at $eB=0.04$ GeV$^2$, 0.08 GeV$^2$ and 0.14 GeV$^2$. For visibility the interpolated results for $N_\tau=8$ and 12 lattices are shown as points.}
    \label{fig:ob_vs_nt_tpc}
\end{figure}

To obtain the continuum estimate based on the available $N_\tau=8$ and 12 lattices, we perform a joint fit using a linear extrapolation in $1/N_{\tau}^2$,
\begin{equation}
    \mathcal{O}\left(T,eB, N_{\tau}\right)=\mathcal{O}(T,eB)+\frac{c}{N_{\tau}^{2}},
    \label{eqn:cont_est_linear}
\end{equation}
where $\mathcal{O}(T,eB)$ is the final continuum estimate. For illustration, we show in~\autoref{fig:ob_vs_nt_tpc} the continuum estimates of $\chi_{11}^{\rm BQ}$ (left) and $-(\mu_{\rm Q}/\mu_{\rm B})_{\rm LO}$ with $n_{\rm Q}/n_{\rm B}=0.4$ along the transition line at three different values of $eB$.

\section{III: Continuum estimate and extrapolation consistency}

In our simulations, we have adopted the HISQ action having the smallest taste symmetry-breaking effects compared to stout and asqtad actions \cite{Bazavov:2011nk}. At vanishing magnetic fields Ref.~\cite{Bollweg:2021vqf} conducted continuum extrapolations based on lattice data for $N_\tau=8,~12,$ and 16 within the HISQ discretization scheme, employing~\autoref{eqn:cont_est_linear}. Here, by utilizing lattice data from Ref.~\cite{Bollweg:2021vqf} based on the linear fit in $1/N_\tau^2$, we affirm the consistency between the continuum estimate and extrapolation for $\chi^{\rm{BQ}}_{11}$ and $-\left(\mu_{\rm{Q}}/\mu_{\rm{B}}\right)_{\rm LO}$ at vanishing magnetic fields. This is demonstrated in Fig. \ref{fig:cont_BQ_q1_eB0}.

\begin{figure}[!ht]
    \centering
    \includegraphics[width=0.39\textwidth]{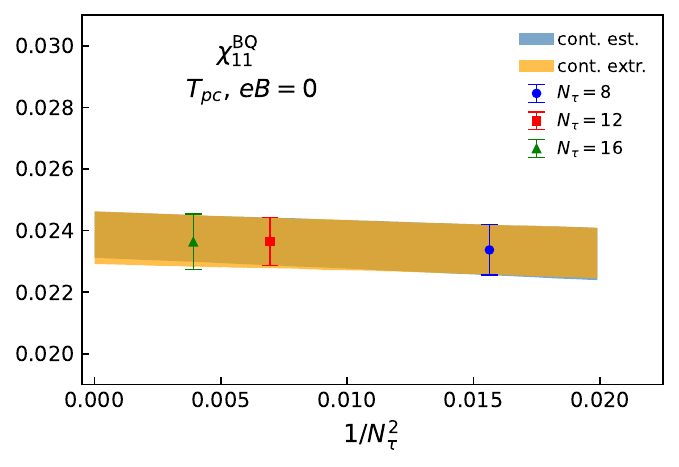}
    \includegraphics[width=0.39\textwidth]{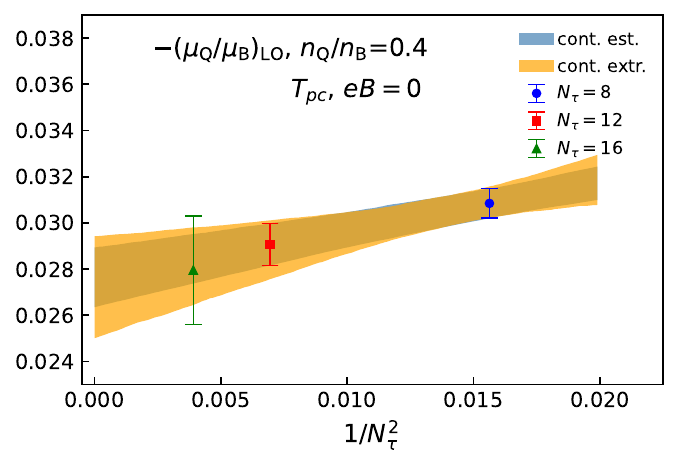}
    \caption{Continuum estimate and extrapolation result of $\chi^{\rm{B} \rm{Q}}_{11}$ (left) and $-\left(\mu_{\rm{Q}}/\mu_{\rm{B}}\right)_{\rm LO}$  with $n_{\rm{Q}}/n_{\rm{B}}=0.4$ along the transition line $T_{pc}$ at $eB=0$ using~\autoref{eqn:cont_est_linear}. The data is taken from Ref.~\cite{Bollweg:2021vqf}.}
    \label{fig:cont_BQ_q1_eB0}
\end{figure}

When extending to a nonzero magnetic field, additional discretization effects arise from the quantization of magnetic flux. To mitigate these effects, it is essential to ensure small magnetic flux in lattice units, that is, $a^2qB\ll 1$, which translates to $N_b/N_\sigma^2\ll 1$. In the literature, $N_b/N_\sigma^2<(2-5)\%$ is commonly adopted \cite{Endrodi:2019zrl, DElia:2021tfb}, while in our investigation, the largest $N_b/N_\sigma^2$ ratio is 0.6\%. Furthermore, we validate the agreement between the continuum estimate and continuum extrapolation at non-vanishing magnetic fields by performing additional lattice QCD computations on $N_\tau=16$ lattices. We generated approximately 3000 configurations at a temperature near the transition region, $T=156.924~{\rm MeV}$, with $N_b = 3$ corresponding to $eB=0.087~{\rm GeV}^2$. Fig. \ref{fig:cont_BQ_q1_eB} shows that at non-vanishing magnetic field, the continuum estimates and continuum extrapolations for $\chi^{\rm{B} \rm{Q}}_{11}$ and $-\left(\mu_{\rm{Q}}/\mu_{\rm{B}}\right)_{\rm LO}$ using~\autoref{eqn:cont_est_linear}, as expected, are consistent with each other, and the systematic uncertainty is within the statistical uncertainties. 	

\begin{figure}[!ht]
    \centering
    \includegraphics[width=0.39\textwidth]{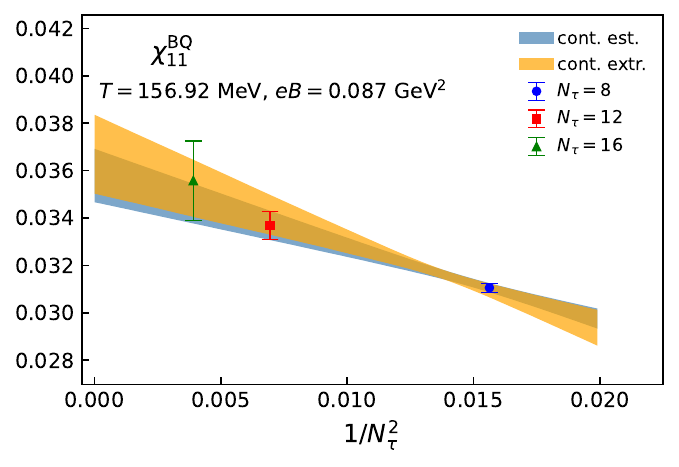}
    \includegraphics[width=0.39\textwidth]{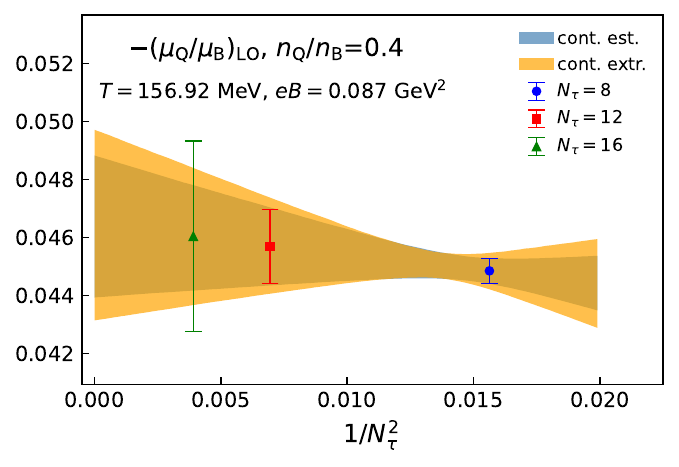}
    
    \caption{Continuum estimate and extrapolation result of $\chi^{\rm{B} \rm{Q}}_{11}$ (left) and $-\left(\mu_{\rm{Q}}/\mu_{\rm{B}}\right)_{\rm LO}$ with $n_{\rm{Q}}/n_{\rm{B}}=0.4$  (right) near the transition temperature $T=156.92$ MeV at $eB=0.087~\text{GeV}^2$. The points for $N_\tau=8$ and 12 are interpolated results while the point for $N_\tau=16$ denotes the lattice data.}
    \label{fig:cont_BQ_q1_eB}
\end{figure}

\section{IV. The construction of proxy for $\chi_{11}^{\rm BQ}$ and $\chi_2^{\rm Q}$ based on the HRG model}
In this section, we show how the proxy for $\chi_{11}^{\rm BQ}$ and $\chi_2^{\rm Q}$ is constructed based on the HRG model following Ref.~\cite{Bellwied:2019pxh}. As being carried out in the experiments, see e.g.~\cite{STAR:2019ans}, net proton number can serve as a proxy for the net baryon number, while the net electric charge is measured through the proton, pion, and kaon.

In order to take into account the contribution from decays of other hadrons into proton ($p$), pion ($\pi$) and kaon ($K$), the proxy for   $\chi_{11}^{\rm B Q}$, i.e. $\sigma_{Q^{\rm PID},p}^{1,1}$, and the proxy for $\chi_{2}^{\rm  Q}$, i.e. $\sigma_{Q^{\rm PID}}^{2}$ are thus constructed in the framework of the HRG following Ref.~\cite{Bellwied:2019pxh} as follows
\begin{equation}
\begin{aligned}
   &\sigma_{Q^{\rm PID},p}^{1,1}= \sum_R\left(P_{R \rightarrow \tilde{p}}\right)\left(P_{R \rightarrow Q^{\rm PID}}\right) I_R^{\rm BQ} + I_{\tilde{p}}^{\rm BQ} ,\\
    &\sigma_{Q^{\rm PID}}^{2}= \sum_R\left(P_{R \rightarrow Q^{\rm PID}}\right)^2 I_R^{\rm Q} + I_{Q^{\rm PID}}^{\rm Q} ,
\end{aligned}
\label{eq:proxy}
\end{equation}
where $P_{R \rightarrow \tilde{j}} = P_{R \rightarrow j} - P_{R \rightarrow \bar{j}}$ indicating the difference between the particle $j$ and its antiparticle $\bar{j}$ with $P_{R \rightarrow j}=\sum_\alpha \mathrm{N}_{R \rightarrow j}^\alpha n_{j, \alpha}^R$ giving the average number of particle $j$ produced by each particle $R$ after the entire decay chain.  $\mathrm{N}_{R \rightarrow j}^\alpha$ represents the branching ratio of decay channel $\alpha$, and $n_{j, \alpha}^R$ is the number of stable particle $j$ produced from this decay channel.  And
$P_{R \rightarrow Q^{\rm PID}}=P_{R \rightarrow \tilde{p}}+P_{R \rightarrow \tilde{K}}+P_{R \rightarrow \tilde{\pi}}$ with $Q^{\rm PID}$ represents the net proton, pion and kaon. In the computation all decay modes of particle $R$ have been taken into account, and the branching ratio is assumed to be independent of the magnetic field, i.e. same as those listed in the Particle Data Group~\cite{ParticleDataGroup:2020ssz}. The $I_R^{\rm BQ}$ and $I_R^{\rm Q}$ stands for the contribution from the particle $R$ to $\chi_{11}^{\rm BQ}$ and $\chi_{2}^{\rm Q}$, respectively, where $\chi_{11}^{\rm BQ}\equiv\sum_i I_{i}^{\rm BQ}$ and $\chi_{2}^{\rm Q}\equiv\sum_i I_{i}^{\rm Q}$ (cf.~\autoref{eq:HRGsus}), while $I_{\tilde{j}}^{X}=I_{j}^{X} - I_{\bar{j}}^{X}$ with $X=\rm BQ,~\rm Q$ and $I_{Q^{\rm PID}}^{\rm Q} =I_{\tilde{p}}^{\rm Q} +I_{\tilde{K}}^{\rm Q}  +I_{\tilde{\pi}}^{\rm Q}$.

\section{V. Next-to-leading order correction to $\mu_{\rm Q}/\mu_{\rm B}$}

The electric charge chemical potential can be expanded as 
\begin{equation}
    \hat{\mu}_{\rm Q}=q_1 \hat{\mu}_{\rm B}+q_3 \hat{\mu}_{\rm B}^3+\mathcal{O}(\hat{\mu}_{\rm B}^5).
\end{equation}
Here the explicit expression of $q_3$ can be found in~\cite{Bazavov:2017dus}, and its evaluation involves computing up to the 4th order fluctuations and correlations of $\rm B$, $\rm Q$, and $\rm S$. To achieve a similar level of precision, the computation of these 4th order fluctuations requires about an order of magnitude more statistics than is required for 2nd order fluctuations. 

To evaluate the next-leading-order correction to $\mu_{\rm Q}/\mu_{\rm B}$, we have performed computations of both 2nd and 4th order fluctuations and correlations on $N_\tau=8$ lattices. The resulting $q_3/q_1$ as a function of $eB$ at $T_{pc}(eB)$ for three different values of $n_{\rm Q}/n_{\rm B}$ is shown in~\autoref{fig:q3_q1_nt8}. It can be seen that $q_3/q_1$ in all cases remains within $\sim$2\%. Similar to the scenario with zero magnetic fields, where no significant discretization effects were noted for $q_3/q_1$~\cite{Bazavov:2012vg,Bazavov:2017dus}, it is thus expected that the next-to-leading order correction to $\mu_{\rm Q}/\mu_{\rm B}$ in the continuum limit will be mild.
\begin{figure}[!ht]    
    \centering           
    \includegraphics[width=0.32\textwidth]{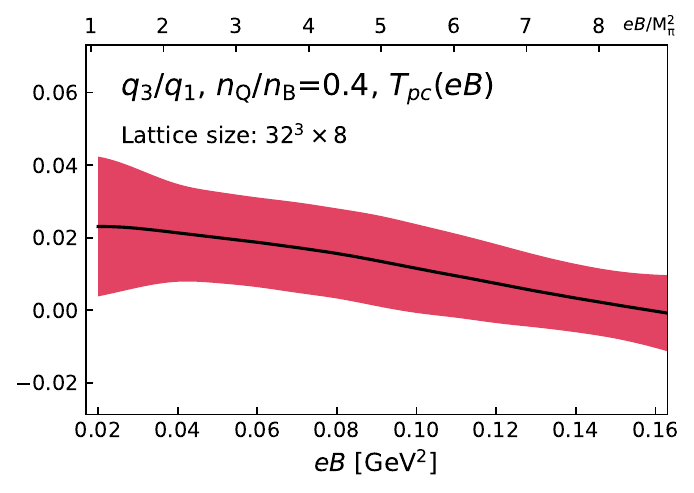}
    \includegraphics[width=0.32\textwidth]{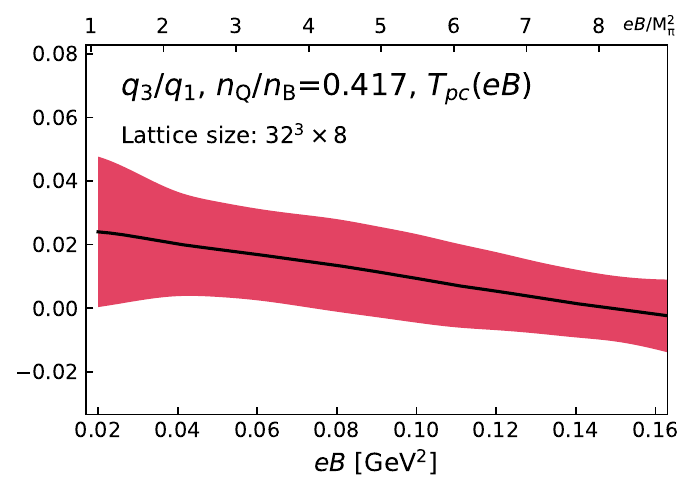}
    \includegraphics[width=0.32\textwidth]{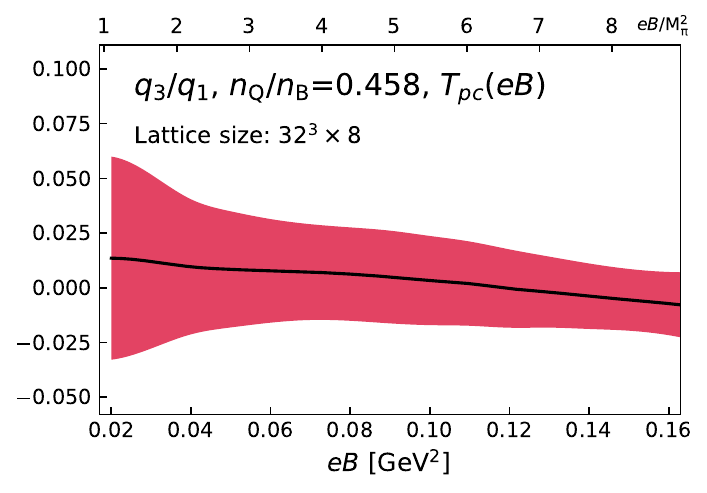}
    \caption{$q_3/q_1$ obtained on $32^3\times 8$ lattices as a function of $eB$ at $T_{pc}(eB)$ for $n_{\rm Q}/n_{\rm B}=0.4$ (left), 0.417 (middle) and 0.458 (right). The value and error of $q_3/q_1$ are denoted by the solid line and band, respectively.} 
    \label{fig:q3_q1_nt8}
\end{figure}

\section{VI: $T_{pc}$ in nonzero magnetic fields}
To determine the crossover transition temperature $T_{pc}$ at nonzero magnetic fields, we investigate the peak location of total chiral susceptibility $\chi_M(eB)$, which is defined as the quark mass derivative of chiral condensate. In our computation, we use the following subtracted chiral condensate $M$ and its corresponding susceptibility $\chi_M$~\cite{HotQCD:2018pds}

\begin{equation}
\begin{aligned}
    &M =\frac{1}{f_K^4}\left[m_s\left(\langle\bar{\psi} \psi\rangle_u+\langle\bar{\psi} \psi\rangle_d\right)-\left(m_u+m_d\right)\langle\bar{\psi} \psi\rangle_s \right]\,, \\
    &\chi_M =\left.m_s\left(\partial_{m_u}+\partial_{m_d}\right) M\right|_{m_u=m_d=m_l} =\frac{1}{f_K^4} \left [m_s\left(m_s \chi_l-2\langle\bar{\psi} \psi\rangle_s-4 m_l \chi_{s u}\right) \right ],
\end{aligned}
\end{equation}
where $\langle\bar{\psi} \psi\rangle_f={T}\left({\partial \ln Z}/{\partial m_f}\right)/{V}$ with $f=u,d,s$ for up, down and strange quark, respectively. Here, $f_K=155.7(9) / \sqrt{2}~ \mathrm{MeV}$ is the kaon decay constant, $\chi_{f g}=\partial_{m_f}\langle\bar{\psi} \psi\rangle_g$, and $\chi_l=\chi_{u u}+\chi_{u d}+\chi_{d u}+\chi_{d d}$. Note that at nonzero magnetic fields, $\chi_{uu}$ and $\chi_{dd}$ are not degenerate, and mixed susceptibilities $\chi_{ud}$ are also taken into account.
\begin{figure}[!htp]    
    \includegraphics[width=0.4\textwidth]{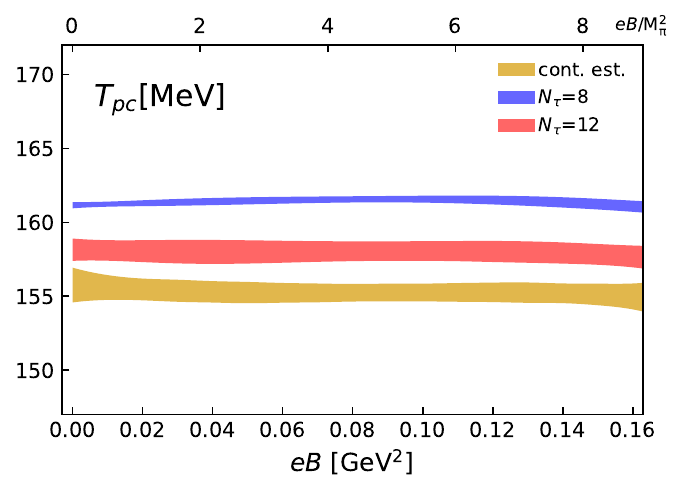}
    \caption{Continuum estimate of $T_{pc}(eB)$.} 
    \label{fig:Tpc}
\end{figure}

Since the magnetic field does not affect the UV-divergent part \cite{Bali:2011qj}, in practice, we determine the peak location of following $\chi_M(eB)$ as $T_{pc}(eB)$
\begin{equation}
    \begin{aligned}
    \chi_M(eB) =\frac{m_s}{f_K^4}\left[m_s \chi_l(eB)-2\langle\bar{\psi} \psi\rangle_s(eB=0)-4 m_l \chi_{s u}(eB=0)\right] .
    \end{aligned}
\end{equation}

The obtained $T_{pc}(eB)$ is shown in~\autoref{fig:Tpc}. It can be seen that $T_{pc}(eB)$ has mild $eB$ dependence in the current window of magnetic field strength.

\section{VII: Supplementary figures for Figs. 2, 3 and 4}

In~\autoref{fig:QBBQatTpc}, we show continuum estimated lattice QCD results and HRG results for $\chi_2^{\rm Q}$ (left), $\chi_2^{\rm B}$ (middle), and $\chi_{11}^{\rm BQ}$ (right) as functions of $eB$ at the transition temperature $T_{pc}(eB)$. These supplementary plots correspond to those presented at $T=145$ MeV in~\autoref{fig:QsusbelowTc}.

\begin{figure*}[!ht]	
\begin{center}
    \includegraphics[width=0.32\textwidth]{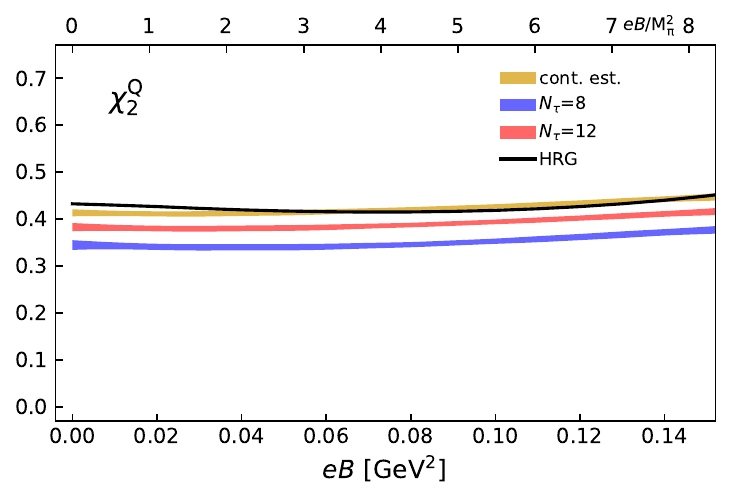}
    \includegraphics[width=0.32\textwidth]{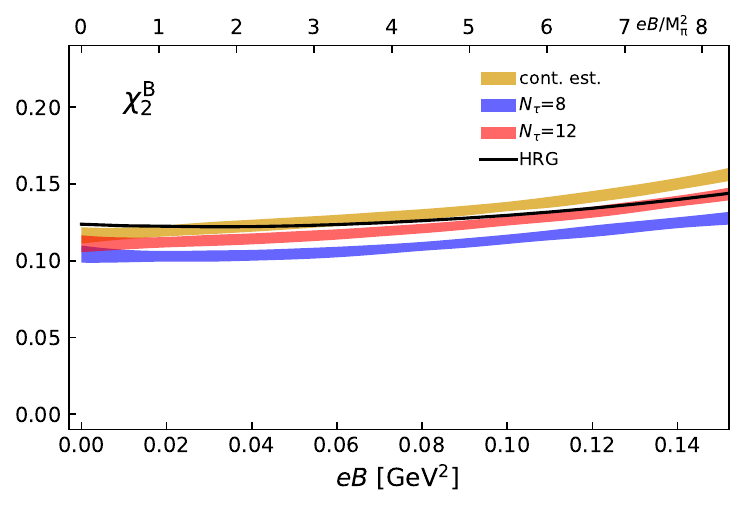} 
    \includegraphics[width=0.32\textwidth]{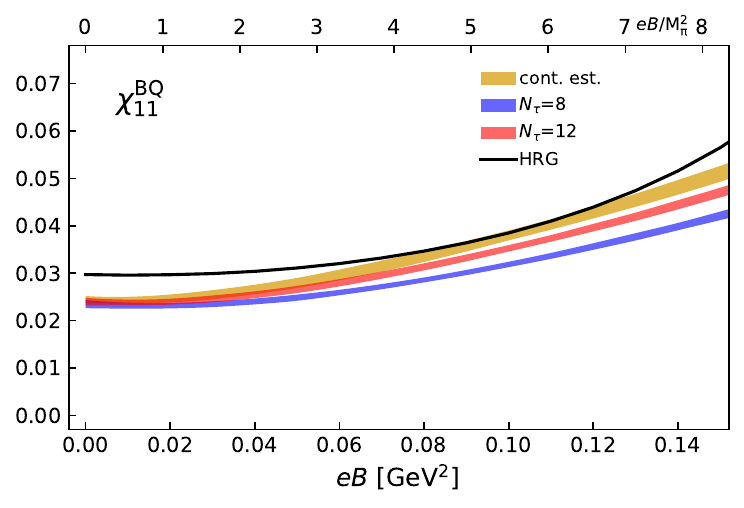}
    \caption{Continuum estimates of  $\chi_2^{\rm Q}$ (left), $\chi_{2}^{\rm B}$ (middle), and $\chi_{11}^{\rm BQ}$ (right) along the transition line based on lattice QCD results with $N_\tau=8$ and $12$. Also shown are results obtained from the  HRG model (black solid lines).}
    \label{fig:QBBQatTpc}		
\end{center}
\end{figure*}

\begin{figure}[h!]			
    \includegraphics[width=0.35\textwidth]{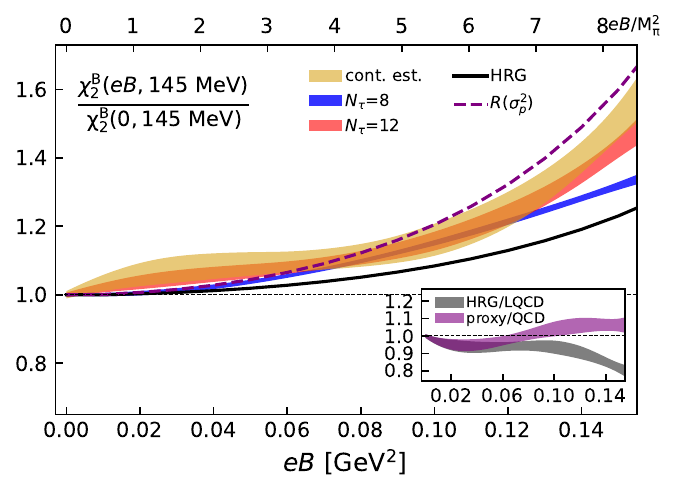} 
    \includegraphics[width=0.35\textwidth]{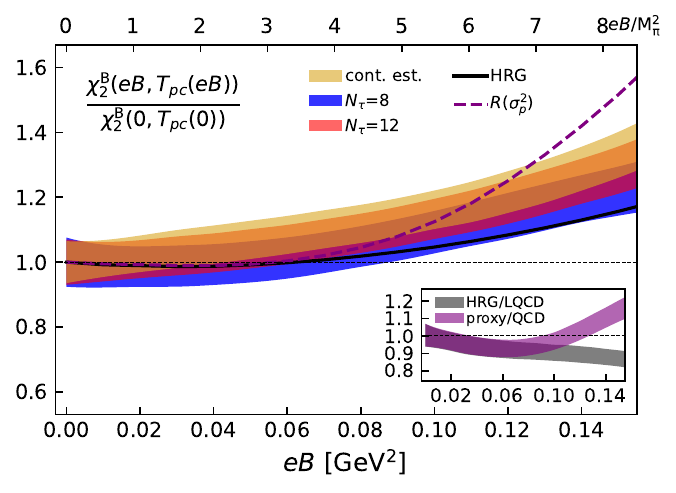} 
    \caption{Ratios of $\chi_{2}^{\rm B}$ to its value at zero magnetic fields at $T=145$ MeV (left) and $T_{pc}(eB)$ (right). The insets show ratios of HRG results and proxy to continuum estimated lattice QCD results. Here $\sigma_p^2$ is constructed in the same way as those in~\autoref{eq:proxy}, i.e. $\sigma_p^2=\sum_R\left(P_{R \rightarrow \tilde{p}}\right)^2 I_R^{\rm B} + I_{\tilde{p}}^{\rm B}$.}
    \label{fig:ratiochi2B}
\end{figure}
In~\autoref{fig:ratiochi2B}, we show $\chi_2^{\rm B}$ normalized to its value at zero magnetic fields as a function of $eB$ at ${\rm T}=145$ MeV (left) and $T_{pc}(eB)$ (right). The corresponding continuum estimated lattice QCD results, HRG results, and the proxy are shown as bands, solid and dashed lines, respectively.

In addition to~\autoref{fig:ratioTpc}, we present  similar results for the normalized $\chi_{11}^{\rm BQ}$ and $\chi_{11}^{\rm BQ}/\chi_2^{\rm Q}$, but at ${\rm T}=145$ MeV in~\autoref{fig:ratiochi11BQQT145}.

\begin{figure}[!ht]			
    \includegraphics[width=0.35\textwidth]{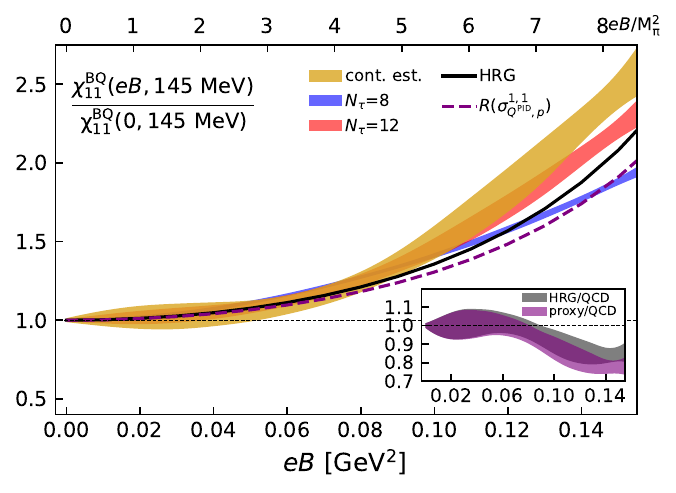}
    \includegraphics[width=0.35\textwidth]{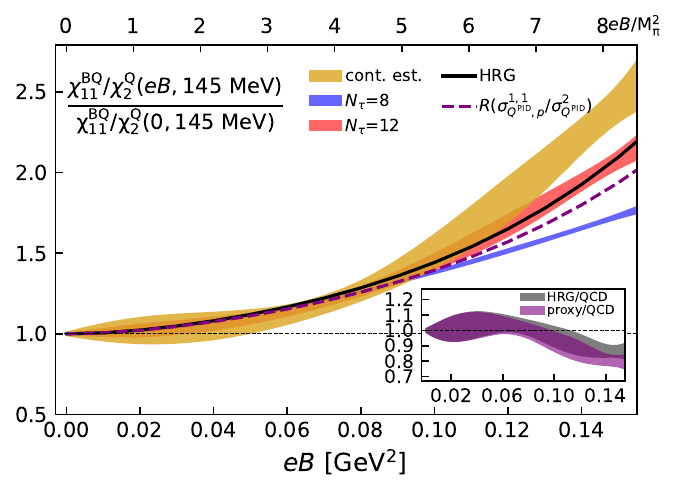}	
    \caption{Ratios of $\chi_{11}^{\rm BQ}$ (left) and $\chi_{11}^{\rm BQ}/\chi_{2}^{\rm Q}$ (right) to their corresponding values at vanishing magnetic fields at $T=145$ MeV. The insets show ratios of HRG results and proxy to continuum estimated lattice QCD results. }
    \label{fig:ratiochi11BQQT145}
\end{figure}

Also, in addition to~\autoref{fig:muQbymuB}, we present similar results of the normalized $(\mu_{\rm Q}/\mu_{\rm B})_{\rm LO}$, but at ${\rm T}=145$ MeV in~\autoref{fig:muQbymuBT145}. Furthermore, we show the unnormalized values, $(\mu_{\rm Q}/\mu_{\rm B})_{\rm LO}$ itself as a function of $eB$ at ${\rm T}=145$ MeV and $T_{pc}(eB)$ in the left and right panels of~\autoref{fig:muQbymuB_value}, respectively.

\begin{figure}[h!]			
    \includegraphics[width=0.35\textwidth]{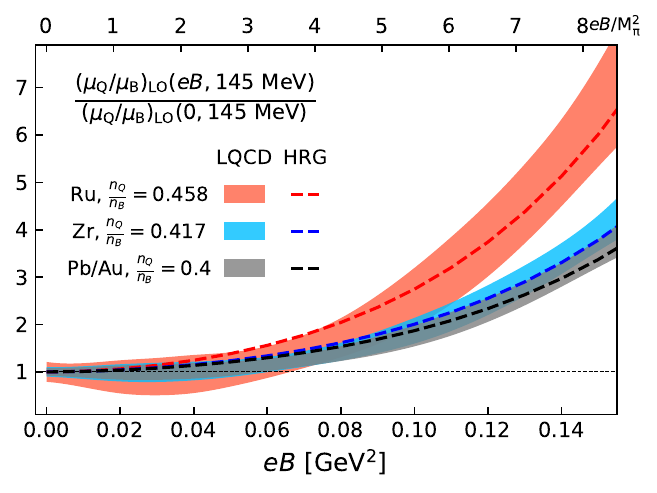} 
    \caption{$(\mu_{\rm Q}/\mu_{\rm B})_{\rm LO}$ normalized to its value at $eB=0$ as a function of $eB$ at $T=145$ MeV.
    Bands correspond to collision systems with various values of $n_{\rm Q}/n_{\rm B}$ and lines are corresponding results obtained from the HRG model.}
    \label{fig:muQbymuBT145}
\end{figure}

\begin{figure}[h!]			
    \includegraphics[width=0.35\textwidth]{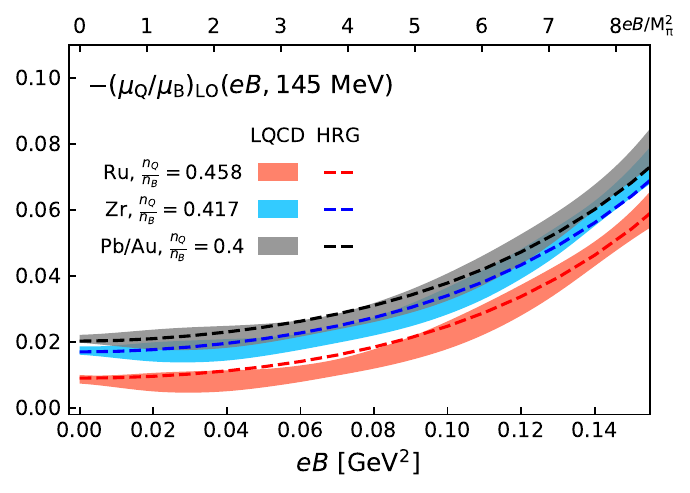} 
    \includegraphics[width=0.35\textwidth]{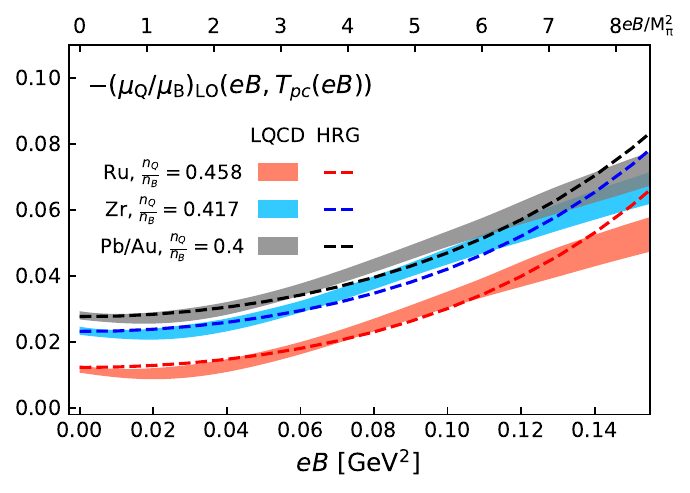} 
    \caption{Continuum estimates of $(\mu_{\rm Q}/\mu_{\rm B})_{\rm LO}$ as a function of $eB$ at $T=145$ MeV (left) and $T_{pc}(eB)$ (right). Bands correspond to collision systems with various values of $n_{\rm Q}/n_{\rm B}$ and lines are corresponding results obtained from the HRG model.}
    \label{fig:muQbymuB_value}
\end{figure}

\end{widetext}

\end{document}